\documentclass[10pt]{article}

\usepackage{times}
\usepackage{graphicx}
\usepackage{amsmath,amssymb}
\usepackage{multirow}
\usepackage{caption}
\usepackage{threeparttable}

\newenvironment{sciabstract}{%
\begin{quote} }
{\end{quote}}

\newcounter{lastnote}

\title{Antenna Design and Implementation for the Future Space Ultra-Long Wavelength Radio Telescope}

\author
{Linjie~Chen,$^{1,2,3\ast}$ Amin~Aminaei,$^{4}$ Leonid~I.~Gurvits,$^{3,5}$ Marc~Klein~Wolt,$^{2}$\\
Hamid~R.~Pourshaghaghi,$^{2,6}$ Yihua~Yan,$^{1}$ Heino~Falcke$^{2,7}$\\
\\
\footnotesize{$^{1}$Key Laboratory of Solar Activity, National Astronomical Observatories, Chinese Academy of Sciences, Beijing 100012, China}\\
\footnotesize{$^{2}$Department of Astrophysics, IMAPP, Radboud University, P.O. Box 9010, 6500 GL Nijmegen, The Netherlands}\\
\footnotesize{$^{3}$Joint Institute for VLBI ERIC, P.O. Box 2, 7990 AA Dwingeloo, The Netherlands}\\
\footnotesize{$^{4}$Department of Physics, University of Oxford, OX1 3RH, United Kingdom}\\
\footnotesize{$^{5}$Department of Astrodynamics and Space Missions, Delft University of Technology, The Netherlands}\\
\footnotesize{$^{6}$Department of Electrical Engineering, Eindhoven University of Technology, 5600 MB Eindhoven, The Netherlands}\\
\footnotesize{$^{7}$Netherlands Institute for Radio Astronomy (ASTRON), P.O. Box 2, 7990 AA Dwingeloo, The Netherlands}\\
\\
\footnotesize{$^\ast$Corresponding author; E-mail:  ljchen@nao.cas.cn.}
}

\date{Feb. 21, 2018}

\begin{document}


\baselineskip18pt


\maketitle


\begin{sciabstract}
In radio astronomy, the Ultra-Long Wavelengths (ULW) regime of longer than 10 m (frequencies below 30 MHz), remains the last virtually unexplored window of the celestial electromagnetic spectrum. The strength of the science case for extending radio astronomy into the ULW window is growing. However, the  opaqueness of the Earth's ionosphere makes ULW observations by ground-based facilities practically impossible. Furthermore, the ULW spectrum is full of anthropogenic radio frequency interference (RFI). The only radical solution for both problems is in placing an ULW astronomy facility in space. We present a concept of a key element of a space-borne ULW array facility, an antenna that addresses radio astronomical specifications. A tripole--type antenna and amplifier are analysed as a solution for ULW implementation. A receiver system with a low power dissipation is discussed as well. The active antenna is optimized to operate at the noise level defined by the celestial emission in the frequency band \(1 - 30\) MHz. Field experiments with a prototype tripole antenna enabled estimates of the system noise temperature. They indicated that the proposed concept meets the requirements of a space-borne ULW array facility.
\end{sciabstract}


\section{Introduction}
\label{intro}The ultra-long wavelength (\(\geq 10~\ m\)) range is presently one of the new promising areas in radio astronomy. The Ultra-Long Wavelength (ULW) band, as the last unexplored regions of the eletromagnetic (EM) spectrum, holds an essential role to understanding of the physical processes in celestial sources. Since the first radio observation in astronomy was performed at this band by Karl Jansky in 1932-33, radio astronomy has moved rapidly to higher frequencies to address the quest for higher spatial resolution and better sensitivity~\cite{diane1997}. However, the ULW band is scientifically very attractive for some key science applications including the origins of the Universe forming of its large structure, evolution of the galaxy, stars, and planetary system, as well as the source of the ultra-high energy cosmic rays~\cite{burns1997},~\cite{heino2009},~\cite{david2011}.

To date, several large Earth-based radio telescopes operating at low frequencies have been built: the Low Frequency Array (LOFAR) in Europe~\cite{haarlem2013}, the Long Wavelength Array (LWA) in New Mexico, USA~\cite{ellingson2009}, and the Murchison Wide-Field Array (MWA) in Western Australia~\cite{tingay2012}. Due to the opaqueness of the Earth's ionosphere, none of these instruments works below 10 MHz. At the frequencies below 30 MHz, the radio astronomy observations are severely limited by strong anthropogenic RFIs. To explore the ULW spectrum region, a spaced-based radio telescope placed beyond the Earth's ionosphere is the only viable solution. A precursor ULW space-borne instrument on-board the Radio Astronomy Explorer (RAE), launched in 1970s ~\cite{herman1973}, discovered that the Moon can act as a good shield for the strong RFI from the Earth to provide an ideal radio environment~\cite{alexander1975}. Measurements conducted by the WIND/WAVES's instrument also revealed that the intensity of terrestrial radio interferences will decrease to an acceptable level~\cite{kaiser1996}. Therefore, in order to have access to the ULW spectral domain with minimum RFIs, the facility must be outside the Earth's ionosphere, either somehow shielded from the Earth-originated RFI, or located sufficiently far away from Earth.

In 1997, ESA has published the report "Very Low Frequency Array on the Lunar Far Side", which is one of the most comprehensive studies on ULW radio observations to date~\cite{bely1997}. Recently, many other concepts have been presented for exploring the ULW region. These range from a single satellite element (DARE~\cite{burn2012} and LRX~\cite{marc2012}) to swarms of small satellites forming multi-element interferometers (FIRST~\cite{bergman2009}, SURO-LC~\cite{baan2013}, OLFAR~\cite{bentum2010},~\cite{raj2015}, DARIS~\cite{boonstra2010}). Their locations cover the lunar surface, lunar orbit, the Sun-Earth L2 point, etc. Within the boundary of a small affordable space mission, a new concept, Discovering the Sky at Longest wavelength(DSL)~\cite{dsl2016}, has been proposed in 2014. It involves one mothership spacecraft and eight nano-satellites placed into a lunar orbit, which enable observations in the RFI-free environment above the far side of the Moon, with the data downlink arranged in the orbital phase above the near side of Moon. DSL would conduct not only single antenna measurements but also form a ULW radio interferometer~\cite{linjie2014}.

In the low frequency domain, a sky noise-limited performance is essential for a radio astronomy antenna system. Since the sky noise at low frequencies is dominant over the noise of the electronics connected with the antenna, even a system with a severe impedance mismatch between antenna and electronics can be acceptable. Accordingly, electrically-short active antennas have been widely employed in HF (3-30MHz) communications for many years~\cite{rhode2001}. As shown in ~\cite{tan2000} this approach could be also applied to low frequency radio astronomy. The antenna design rules and performance boundaries at low frequencies within this approach has been fully discussed~\cite{ellingson2005}.

To observe the ULW radio sky emission, the RAE mission employed single long dipole antenna operating at several fixed frequencies~\cite{herman1973}, which is very limited in resolution and sensitivity. In the ESA report~\cite{bely1997} a 4 m active dipole antenna was suggested as a receiving element for the ULW array on the lunar far side. In other recent studies~\cite{bergman2009},~\cite{baan2013},~\cite{burn2012},~\cite{bentum2010},~\cite{marc2012}, dipole antennas have also been considered for ULW astronomy. For the ULW antenna design, discussed different antenna concepts operating from 1 to 10 MHz have been discussed in 2007 with the conclusion that a non-matched dipole is acceptable for the space-borne radio astronomy~\cite{arts2010}. An active antenna prototype for the Lunar Radio eXplorer (LRX) for the operating frequency band 1 to 100 MHz has been developed and tested~\cite{linjie2011}. However, so far, no practical implementations under scientific and technology constraints of these approaches has been accomplished yet.

This paper presents a design of an active antenna for the ULW radio telescope in the DSL project. Its scientific objectives which define the requirements to the antenna system are briefly described in Section 2. In Section 3, the system noise model is analyzed to establish some limitations for the antenna design. In Section 4 the design of an active antenna with its low noise amplifier (LNA), and the overall system architecture is presented. In Section 5 results of field tests of the developed prototype antenna are discussed. Conclusions are summarized in Section 6.

\section{Antenna system requirements}
\label{sec:requ} At low frequencies, synchrotron emission is often the main mechanism of radiation, where plasma absorption, dispersion and refraction plays important roles in radio wave propagation. According to the modern paradigm, the ULW emission is generated in a broad variety of astrophysical objects ranging from galaxies and
their clusters, active galactic nuclei, stellar and planetary systems, interstellar medius. Cosmic rays of all energies also have an ULW "footprint". Especially attractive is a prospect of observing a red-shifted HI 21 cm spectrum lines from the cosmological dark ages and cosmic dawn~\cite{heino2009}.

\begin{table}[ht]
\caption{Science requirements; I is intensity, FS denotes full
stokes~\cite{dsl2016}.} \label{tab:scireq}
\renewcommand{\multirowsetup}{\centering}
\begin{tabular}{c|c|c|c|c|c|c}
\hline & \multirow{3}{2.0cm}{Dark Ages} &
\multirow{3}{1.0cm}{Galactic discrete sources} &
\multirow{3}{1.1cm}{Galactic diffuse emissions}&
\multirow{3}{1.0cm}{Extra-galactic sources}&
\multirow{3}{1.5cm}{Solar-terr. physics, transient}&
\multirow{3}{1.2cm}{Generalized requirement}\\
&&&&&&\\
&&&&&&\\
\hline
 \multirow{3}{1.2cm}{Science products}& global sky &\multicolumn{3}{c|}{}&\multirow{3}{1.5cm}{triggered observation FoV}&\\
&signal&\multicolumn{3}{c|} {all-sky map}&&all\\
 &spectra&\multicolumn{3}{c|}{}&&\\
\hline
 Freq. Range &&\multicolumn{3}{c|}{}&&\\
 (MHz)&&\multicolumn{3}{c|}{}&&\\
 \emph{main cases} &10 -- 30&\multicolumn{3}{c|}{1 -- 30}&0.5 -- 30&0.5 -- 30\\
 \emph{cross ref.} &10 -- \(\geq50\)&\multicolumn{3}{c|}{1 -- \(\geq50\)}&0.1 -- \(\geq50\)&0.1 -- \(\geq 50\)\\
\hline
 Instant.&\multirow{3}{1.2cm}{30, 1 for calibration}&\multicolumn{3}{c|}{\multirow{3}{*}{1}}&\multirow{3}{*}{\(\leq30\)}&\multirow{3}{*}{\(\leq30\)}\\
 Bandwidth&&\multicolumn{3}{c|}{}&&\\
 (MHz)&&\multicolumn{3}{c|}{}&&\\
\hline
 Spectral &\multirow{3}{1.2cm}{\(10^{3}\), 1 for calibration}&\multicolumn{3}{c|}{\multirow{3}{*}{\(10^{3}\)}}&\multirow{3}{*}{0.1 -- 10}&\multirow{3}{*}{0.1 -- \(10^{3}\)}\\
 resolution &&\multicolumn{3}{c|}{}&&\\
 (kHz) &&\multicolumn{3}{c|}{}&&\\
\hline
 Spatial &&\multicolumn{3}{c|}{}&& \\
 \emph{FoV} &\(4\pi~sr\)&\multicolumn{3}{c|}{\(4\pi~sr\)}&\(4\pi~sr\)&\(4\pi~sr\)\\
 \emph{resolution} &\(4\pi~sr\)&\multicolumn{3}{c|}{\(10^{\prime}\)@ 1 MHz, \(1^{\prime}\)@ 10 MHz}&\(1^{\prime}\) -- \(5^{\prime}\)&1' -- \(4\pi~sr\)\\
\hline
 \multirow{2}{2.0cm}{Sensitivity} &\multirow{2}{2.0cm}{5 mK in 8 months}&\multicolumn{3}{c|}{\multirow{2}{3.1cm}{100 mJy@ 10 MHz and 5.6 Jy@ 1 MHz in a yr}}&\multirow{2}{*}{varying}&\multirow{2}{*}{varying}\\
 &&\multicolumn{3}{c|}{}&&\\
\hline
\multirow{2}{2.0cm}{Temporal Resolution} &\multirow{2}{2.0cm}{1 min -- 8 months}&\multicolumn{3}{c|}{\multirow{2}{3.1cm}{1 month}}&\multirow{2}{*}{0.01 -- 60s}&\multirow{2}{*}{varying}\\
 &&\multicolumn{3}{c|}{}&&\\
\hline
\multirow{2}{2.0cm}{Polarization} &\multirow{2}{2.0cm}{I, FS for calibration}&\multirow{2}{*}{I, FS}&\multirow{2}{*}{I}&\multirow{2}{*}{I, FS}&\multirow{2}{*}{FS}&\multirow{2}{*}{FS}\\
 &&&&&&\\
\hline
\end{tabular}
\begin{tablenotes}
    \item (1) Due to lunar sky-blocking, instantaneous \(FoV < 4\pi~sr\);
    \item (2) \(10^{6}\)Jy is a typical sensitivity level at 15 MHz with a bandwidth of
    \(10^{3}\) kHz for a normal solar burst.
    \item (3) I is intensity, FS denotes full Stokes.
   \end{tablenotes}
\end{table}

Discovering the Sky at the Longest Wavelengths (DSL) is a small-scaled mission presented by a joint China-Europe scientific team~\cite{dsl2016}. It will consist of a mother ship and eight daughter-satellites in near-identical low altitude lunar orbits. These satellites form a quasi-linear array, quickly filling the observational aperture in each orbit. Orbit precession ensures filling of an entire three dimensional aperture, allowing all-sky observations. Table~\ref{tab:scireq} summarizes the requirements to the radio astronomy instrumentation of the DSL mission for various science tasks.

The frequency band \(1 - 30\) MHz is chosen as the main one for addressing the science applications in Tab.~\ref{tab:scireq} beyond reach for the Earth-based instruments. For additional science tasks and cross-verification with Earth-based and space-borne facilities such as LOFAR and WIND/WAVES, the frequency band can be extended up to 50 MHz and down to about 0.1 MHz respectively. The instantaneous bandwidth requirement varies over the range of the science applications. For synthesis imaging, the total available bandwidth is limited to 1 MHz by the maximum inter-satellite data rate of 2 Mega bits per second (Mbps) for communication. To carry out synthesis imaging with high spectral resolution, the sky noise-limited performance is essential for the antenna system. The total payload mass and power budget for DSL radio telescope are limited to 60 kilograms (kg) and 65 Watts (W) respectively, we define the mass and power requirements for antenna system on each satellite as 2 kg and 4 W respectively. To observe the full polarization, sets of two or three colocated (quasi-)orthogonal dipoles or monopoles are required~\cite{zarka2012}.

Table~\ref{tab:antreq} defines the technical requirements for the ULW onboard antenna system.

\begin{table}[h]
\caption{Antenna system requirements} \label{tab:antreq}
\centering
\begin{tabular}{|c|c|}
\hline

Frequency range & \(1-30\) MHz \\

Instantaneous bandwidth & \(\geq1\) MHz \\

Spectral resolution & \(\geq 1\) kHz \\

System noise level& \(N_{rec} < 10\% N_{sky}\)\\

Number of independent polarization measures  & \(\geq 2\) \\

Mass & \(< 2\) kg \\

Power consumption & \(< 4\) W \\

\hline
\end{tabular}
\vspace{-7pt}\renewcommand{\footnoterule}{}
\end{table}

\section{System Noise Analysis}
\label{sec:temmodel} 
\begin{figure}[th]
\centering
\setlength{\belowcaptionskip}{-0.5cm}
\includegraphics[width=3.5in]{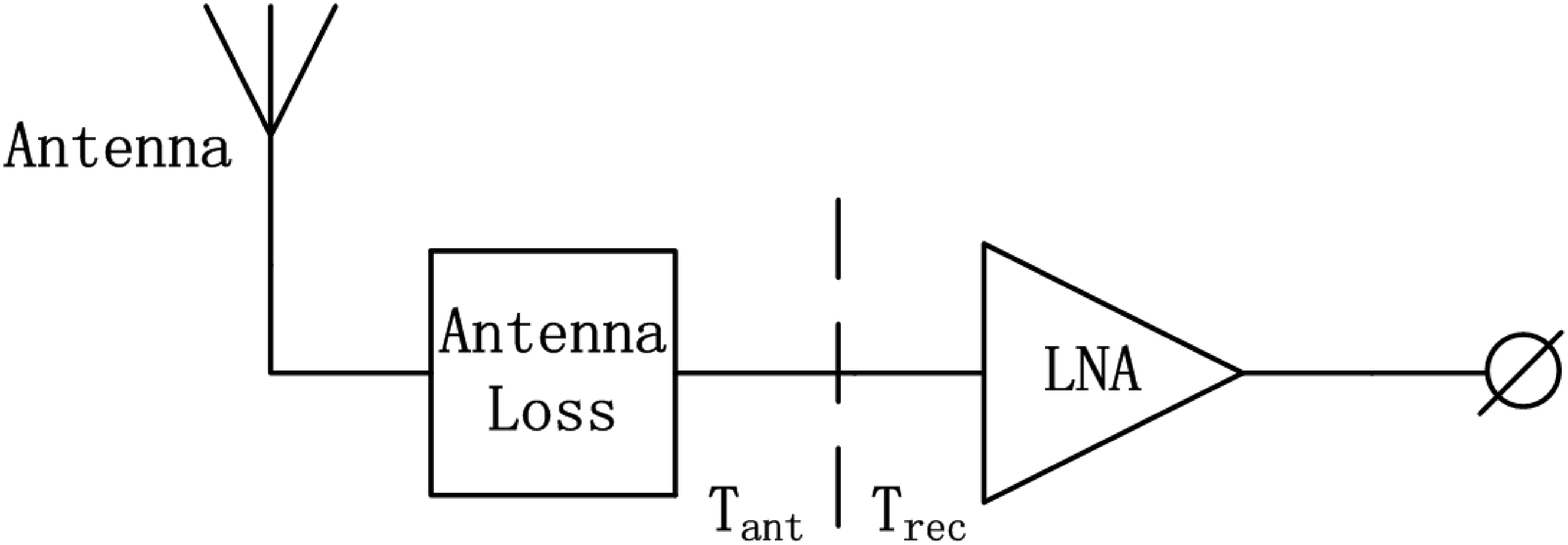}%
\caption{Noise model of an antenna with preamplifier.
\label{fig:antmodel}}
\end{figure}

The single antenna system of a low frequency radio telescope can be modeled as antenna, low noise amplifier (LNA) and receiver (without LNA) shown in Fig.~\ref{fig:antmodel}. To simplify the noise analysis of the antenna system, we assume the receiver noise being negligibly low in Fig.~\ref{fig:antmodel} since the LNA noise is dominant over the noise of the electronics of the receiver. The noise induced on the antenna includes the thermal noise of the antenna itself and the noise from the sky. Due to the ohmic losses of the antenna, its efficiency \(\eta\) will always be less than 1. At the reference plane between antenna and LNA input, the relation between the antenna noise temperature \(T_{ant}\), the sky temperature \(T_{sky}\) and the ambient temperature \(T_{0}\) (assumed to be 290 K) is given by
\begin{equation}
\label{eqn:antnoise}
T_{ant}=\eta T_{sky}+(1-\eta) T_{0}
\end{equation}

%
Assuming the sky background noise is uniform over the celestial hemisphere, it can be then approximated as~\cite{heino2009},
\begin{eqnarray}
\hspace{-8mm} T_{sky} \!=\! \label{eqn:skynoise}
\begin{cases}
16.3\!\times\! 10^{6}(\displaystyle\frac{\nu}{\nu_{0}})^{-2.53}K & \nu \!>\! 2 \text{ MHz},\\[10pt]
16.3\!\times\! 10^{6}(\displaystyle\frac{\nu}{\nu_{0}})^{-0.3}K &
\nu \!\leq\! 2 \text{ MHz}.
\end{cases}
\end{eqnarray}

where \(\nu_{0}=2\) MHz. The LNA noise temperature is defined at the reference plane to show the noise level of the LNA itself, which is frequency dependent. Letting \(S_{ant}\) be the available power spectral density of the antenna noise due to \(T_{ant}\) at the output of LNA, one finds

\begin{equation}
S_{ant}=kT_{ant}(1-|\Gamma_{ant}|^{2})G_{amp} \label{eqn:santt}
\end{equation}

where \(\Gamma_{ant}\) is the voltage reflection coefficient with the input impedance of the LNA as reference impedance, \((1-|\Gamma_{ant}|^{2})G_{amp}\) denotes the available gain of the LNA. Substituting \(T_{ant}\) with (\ref{eqn:antnoise}), the above equation can be rewritten as

\begin{eqnarray}
\hspace{-8mm}S_{ant}=k\eta T_{sky}(1-|\Gamma_{ant}|^{2})G_{amp}+k(1-\eta)T_{0}(1-|\Gamma_{ant}|^{2})G_{amp} \label{eqn:sant}
\end{eqnarray}

The first term describes the power spectral density due to the sky noise, and the second term describes the antenna thermal noise power density. Finally, the noise power density arising from the LNA noise being a function of \(\Gamma_{ant}\) at the output of LNA is

\begin{equation}
\emph{S}_{\emph{LNA}}=\emph{kT}_{\emph{LNA}}(1-|\Gamma_{\emph{ant}}|^{2})\emph{G}_{\emph{amp}}
\label{eqn:slna}
\end{equation}

where \(T_{LNA}\) is the LNA's noise temperature. The sky noise-limited performance of antenna system requires

\begin{equation}
S_{LNA}+S_{ant,th} < \frac{1}{10}S_{sky} \label{eqn:skl}
\end{equation}

After substituting (\ref{eqn:sant}) and (\ref{eqn:slna}) in (\ref{eqn:skl}), the constraint of the LNA noise is given as

\begin{equation}
\emph{T}_{\emph{LNA}} < \frac{1}{10}\eta \emph{T}_{\emph{sky}}-(1-\eta)\emph{T}_{0} \label{eqn:tlnaui}
\end{equation}

For a non-matched system as described above, the LNA noise becomes very high due to the bad mismatching between the source and the LNA, which is not comparable with the matched LNA, and hard to determine if this noise temperature is realistic~\cite{arts2010}. To better understand the LNA noise performance, we introduce an equivalent noise temperature \(T^{\prime}_{LNA}\) of a matched system. The output noise power density of the equivalent matched LNA is \(S^{\prime}_{LNA}=kT^{\prime}_{LNA}G_{amp}\) . Let the output noise of the matched LNA \(S^{\prime}_{LNA}\) equals to the output noise \(S_{LNA}\) of the non-matched LNA. Substituting \(S_{LNA}\) with (\ref{eqn:slna}), we get

\begin{equation}
\emph{T}^{\prime}_{\emph{LNA}}=(1-|\Gamma_{\emph{ant}}|^{2})\emph{T}_{\emph{LNA}}
\label{eqn:slnam}
\end{equation}

After substituting (\ref{eqn:tlnaui}) in (\ref{eqn:slnam}), the constraint of the equivalent LNA noise in a matched system is given as

\begin{equation}
\emph{T}^{\prime}_{\emph{LNA}} < (1-|\Gamma_{\emph{ant}}|^{2})[\frac{1}{10}\eta \emph{T}_{\emph{sky}}-(1-\eta)\emph{T}_{0}] \label{eqn:tlnaui1}
\end{equation}

The impedance match between the antenna and LNA is often characterized by the voltage standing wave ratio (VSWR), which is defined as \((1+|\Gamma|)/(1-|\Gamma|)\), therefore (\ref{eqn:tlnaui1}) can be written as

\begin{eqnarray}
\hspace{-8mm}\emph{T}^{\prime}_{\emph{LNA}} \!<\!
\frac{4\cdot\footnotesize\text{VSWR}}{(\footnotesize\text{VSWR}+1)^{2}}[\frac{1}{10}\eta T_{sky}\!-\!(1\!-\!\eta) T_{0}]
\label{eqn:tlnavswr}
\end{eqnarray}

For the ideally matched case (\(\footnotesize\text{VSWR} \approx 1\)), the formula (\ref{eqn:tlnavswr}) reduces to (\ref{eqn:tlnaui}), and for the extremely badly matched case (\(\footnotesize\text{VSWR} \gg 1\)),
\begin{equation}
\emph{T}^{\prime}_{\emph{LNA}}<\frac{4}{\footnotesize\text{VSWR}}[\frac{1}{10}\eta T_{sky}\!-\!(1\!-\!\eta) T_{0}]
\label{eqn:tlnanm}
\end{equation}

\begin{figure}[t]
\centering
\setlength{\belowcaptionskip}{-0.5cm}
\includegraphics[width=\columnwidth]{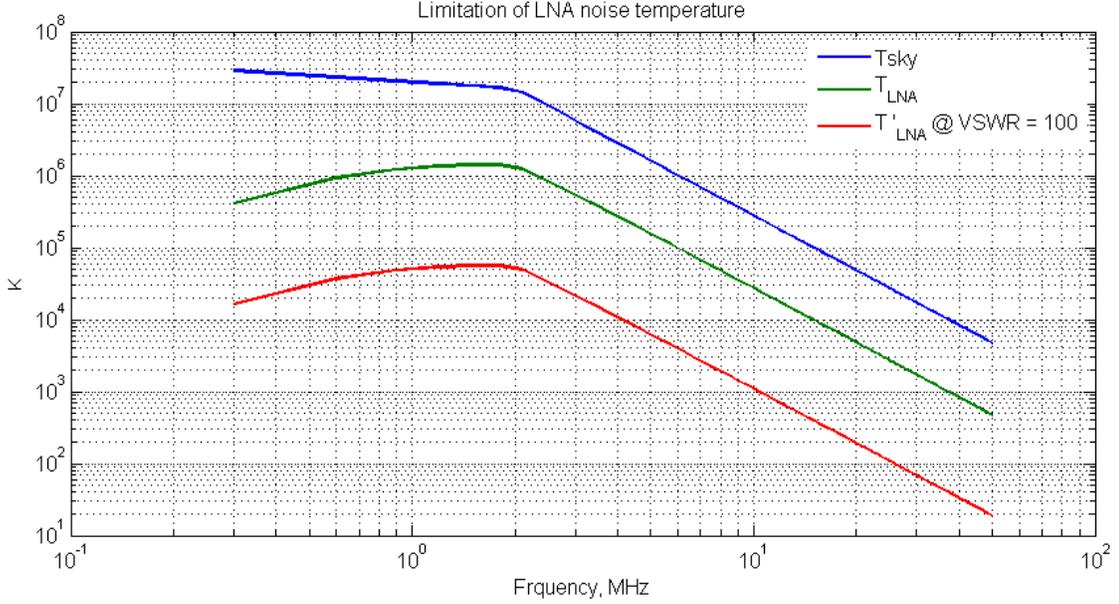}%
\caption{LNA noise temperature limitation for a 5-meter dipole antenna in space over frequency, \(T_{0} = 290\) K. The blue line is the sky temperature, the green one shows the required LNA noise temperature for a non-matched system, and the red one represents the required LNA noise temperature defined in a matched system. \label{fig:reqnoise}}
\end{figure}

In Fig.~\ref{fig:reqnoise}, the noise constraints of LNA are plotted under requirement of achieving the sky noise-limited performance. It can be seen that the required LNA noise is above \(10^{3}\) K across all the operating frequency band, which is easy to meet. For the LNA in a matched system defined above, although in practice an antenna does not have a constant VSWR over operating frequencies, this plot is still useful for understanding how good the LNA should be to obtain the sky noise-limited operation. It can be seen that the required LNA noise is above \(10^{3}\) K below 10 MHz and no less than \(10^{2}\) K between 10 MHz and 30 MHz  even if the antenna is badly matched. These requirements are not difficult to meet technically.

\section{Antenna System Designs}
\label{sec:design}
Many broadband antennas have already proposed to realize a large operational frequency band at long wavelenghts~\cite{hicks2012}\cite{gemmeke2009}. But taking into account mechanical design complexities of such antennas, they are not suitable for a space mission. Dipole antennas are also widely used in communication systems, same as magnetic loop antenna~\cite{eisenberg1962}\cite{tiehan1991}. However, magnetic loop antennas are impractical in the low frequency domain in space due to their large dimensions to meet the noise-limited performance, and matched dipoles are unfeasible either since they can hardly achieve the required instantaneous bandwidth~\cite{arts2010}. Although non-matched dipoles have low overall efficiency, they are still able to operate at the sky noise-limited operation over a broad frequency band, taking into account the dominant sky noise at low frequency. For the DSL space radio telescope, a non-matched antenna combined with low noise amplifier has been selected as the receiving element.

In this section, we present the antenna system designs including an antenna element, low noise amplifier and on-board receiver.

\begin{figure}[ht]
\centering
\setlength{\belowcaptionskip}{-0.5cm}
\includegraphics[width=3.5in]{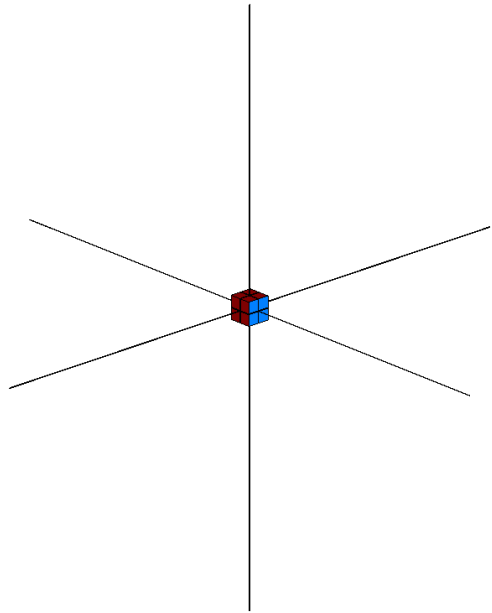}%
\caption{Scale model of a tripole antenna (5.0 m) on a
micro-satellite (\(30\times30\times30\)~cm). \label{fig:tripole}}
\end{figure}

\subsection{The Antenna}
\label{subsec:antenna}

At low frequency the suitability of a dipole antenna for a space borne application is primarily in its simplicity, a dipole antenna is simple and easy to be deployed for a space mission. In principle, two cross-dipole antennas are sufficient to obtain all the polarization properties of an incoming wave. However, adding another orthogonal dipole will resolve well the sensitivity lack in specific directions for the cross dipoles, and will increase the detection capability, this is a triple dipole antenna, and also called tripole antenna~\cite{compton1981} (Fig.~\ref{fig:tripole}). A tripole antenna is sensitive to a three-dimension electric field, which is quite helpful to increase the field of view of radio telescope in mapping the radio sky. It can also protect desired signals from almost any interference signals~\cite{compton1981}, which is considered as a big advantage given the fact that interference is a fatal threat at low frequencies. Without spatially distributed arrays a single tripole antenna is capable of estimating the directions of arrival (DOA) of incident signals~\cite{linjie2010}, which enables us to localize discrete sources including transient radio sources. All these aforementioned properties make the tripole antenna very suitable for space-based radio observations.

Due to the limitation of the mass and dimension, it is unfeasible to mount a resonant antenna to receive the radio signals on a micro-satellite, especially at very low frequencies. The antenna should be short comparing to the wavelength, and this implies that its impedance is capacitive. At the output of the antenna, the voltage of the induced signal is dependent on the radio emission intensity. A combination of the antenna and pre-amplifier is usually called an active antenna, which has been widely used in the modern low frequency radio array systems in the ground-based facilities~\cite{hicks2012}\cite{tan2000}.

\begin{figure}[t]
\centering
\setlength{\belowcaptionskip}{-0.5cm}
\includegraphics[width=\columnwidth]{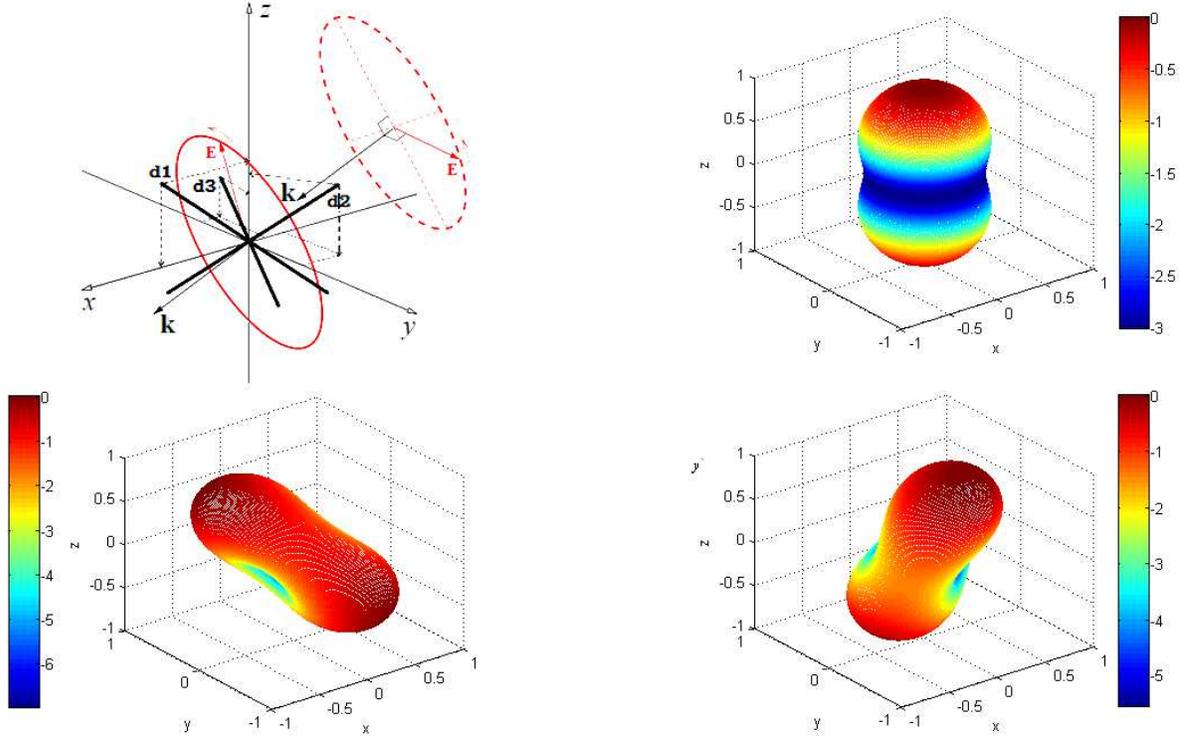}%
\caption{Radiation patterns of a tripole antenna in a coordinate system shown in upper-left panel for different excitations. The three dipoles of this tripole antenna are placed symmetrically around z axis, the angles between all three dipoles and xy plane are the same, \(35.3^{\circ}\). In this case, all these three dipoles are mutually orthogonal and have the same conditions with respect to xy plane. \(Upper \ right:\ A_{1}=1, A_{2}=1, A_{3}=1. \theta_{1}=0, \theta_{2}=2\pi/3, \theta_{3}=-2\pi/3.\ Bottom \ left:\ A_{1}=2, A_{2}=0, A_{3}=1. \theta_{1}=\pi/6, \theta_{2}=0, \theta_{3}=-\pi/3.\ Bottom \ right:\ A_{1}=1, A_{2}=2, A_{3}=1. \theta_{1}=0, \theta_{2}=2\pi/3, \theta_{3}=-3\pi/4.\)Here \(A_{1}, A_{2}, A_{3}\ and\ \theta_{1}, \theta_{2}, \theta_{3}\) indicate respectively the amplitudes and phases of the excitations applied on the three dipoles d1, d2 and d3. The radiation patterns are scaled in dB~\cite{linjie2011}.
\label{fig:pattern}}
\end{figure}

The Tripole antenna consists of three mutually orthogonal dipoles, and can be considered as a co-centered antenna array. Similar to the beam forming concept, we can multiply the induced signal on each dipole of the tripole antenna with a complex weight, and then combine them together to get a desired beam~\cite{eriksson2003}. As shown in Fig.~\ref{fig:pattern}, this property is quite useful for RFI suppression or some scientific experiments that just concern the sky radio sources in specific directions.

To evaluate the performance of the proposed active tripole antenna, we simulate the antenna onboard a micro-satellite modeled as a cube with the dimensions of \(30\times30\times30\) cm~\cite{dsl2016}. The three dipoles are modeled as copper strips with a width of 8 mm and a thickness of 0.12 mm. They are mounted in the three orthogonal directions of the satellite cube. The dipole length ranges from 5 meters to 15 meters.

\begin{figure}[t]
\centering
\setlength{\belowcaptionskip}{-0.5cm}
\includegraphics[width=\columnwidth]{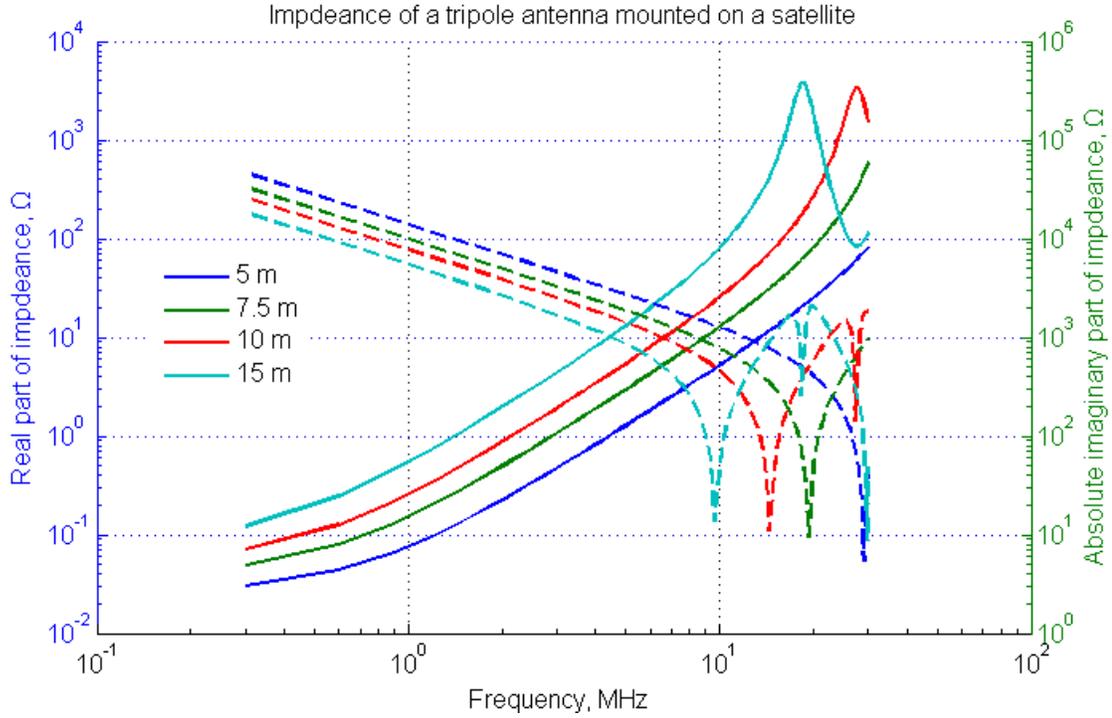}%
\caption{Impedance of a tripole antenna onboard a micro-satellite. The solid lines and dashed lines represent real and imaginary parts,
respectively.  \label{fig:impedance}}
\end{figure}

Figure~\ref{fig:impedance} shows the simulated impedance of a single dipole of the proposed tripole antenna. The real impedance ranges from about \(10^{-1} \rm \Omega \) to about \(10^{3} \rm\Omega\) over the frequency range, and the imaginary impedance varies between about \(10^{0} \rm\Omega\) and about \(10^{4}\rm\Omega\). It also shows that the resonance frequency of the antenna decreases with the increasing of its length, as expected (Note that all the antenna simulation results are produced by the software Zeland IE3D in this work).

To achieve the frequency range from 1 to 30 MHz, it is supposed to set the resonance frequency at 15 MHz for the antenna. However, at low frequencies the sky noise is dominant, it increases with power law towards the low end of the frequencies. This will inevitably increase the system temperature and thus reduce the sensitivity of the antenna. For some specified science experiments, such as dark age and transient, radio detection with extremely high sensitivity is required, therefore, it is necessary to optimize the resonance frequency to achieve the best antenna sensitivity. Here we choose to increase the resonance frequency, which reduces the antenna efficiency at the low end of the frequency band. Since the sky noise dominates the receiving noise, according to formula (\ref{eqn:antnoise}), the antenna noise temperature will clearly decrease, and thus the sensitivity will be improved. As a result, the resonant frequency of the antenna is designed around 20MHz with a trade-off between the sensitivity and the frequency band, which corresponds to the antenna length of 7.5 meters. The simulations of the sky noise-limited performance in Subsec.~\ref{subsec:snl} also proved that 7.5 meters is a feasible length for the tripole antenna.

\subsection{The Low Noise Amplifier}

\begin{figure}[t]
\centering
\setlength{\belowcaptionskip}{-0.5cm}
\includegraphics[width=\columnwidth]{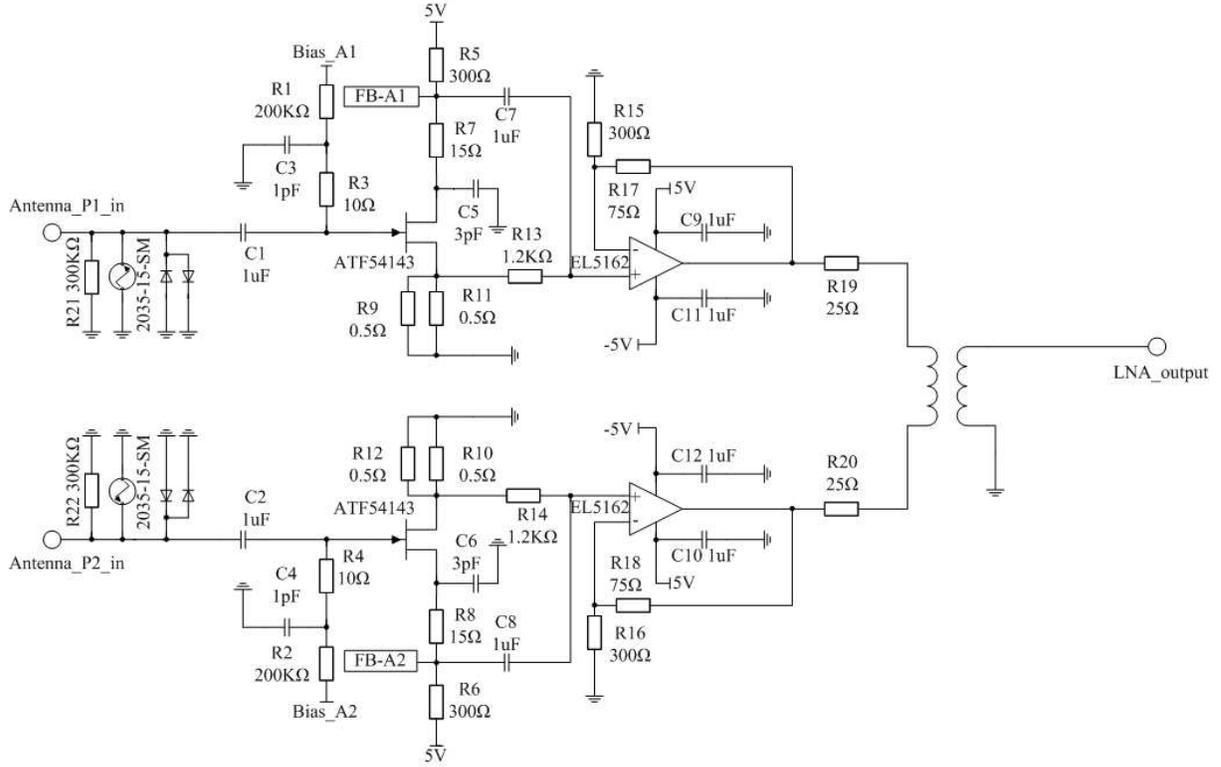}%
\caption{Schematic of the low noise amplifier. The blocks FB-A1 and FB-A2 denote the feedback circuits of the first stage. Here the Electro-Static discharge (ESD) protection circuits are designed in the front of the first stage to efficiently decrease the voltage that can damage the E-PHEMT while keeping the input capacitance low in respect to the antenna capacitance. \label{fig:LNA}}
\end{figure}

\label{subsec:amplifier}The low noise amplifier is directly connected with the terminal of the antenna. As a vital component of an active antenna, it amplifies the induced signal on the antenna without introducing much self-noise, meanwhile, it serve as a balun to achieve the impedance matching and the signal transfer between
balance and unbalance signals.

In Fig.~\ref{fig:impedance} it can be seen that the variation of the tripole antenna impedance is extremely large, which makes it very difficult to achieve either power or noise matching between the LNA and the antenna within such broad frequency range. A solution is in the design of a LNA with a very low input impedance (current amplifier), or a very high input impedance (voltage amplifier). An antenna with a current amplifier has a high Q (quality factor)resonance peak in its transfer response. For the proposed tripole antenna, the resonance peak is at around 20 MHz. The disadvantage of a high Q is that the transfer response becomes very sensitive to the antenna environment, which results in more vigorous calibrations than that of a voltage amplifier. In addition, coupling between two antennas close to each other depends on the input impedance of the LNA. Considering there are many other electronic circuits on the micro-satellite, the coupling should be avoided as much as possible. Therefore, a voltage amplifier has been designed as the LNA for the active tripole antenna.

\begin{figure}[h]
\centering
\setlength{\belowcaptionskip}{-0.5cm}
\includegraphics[width=\columnwidth]{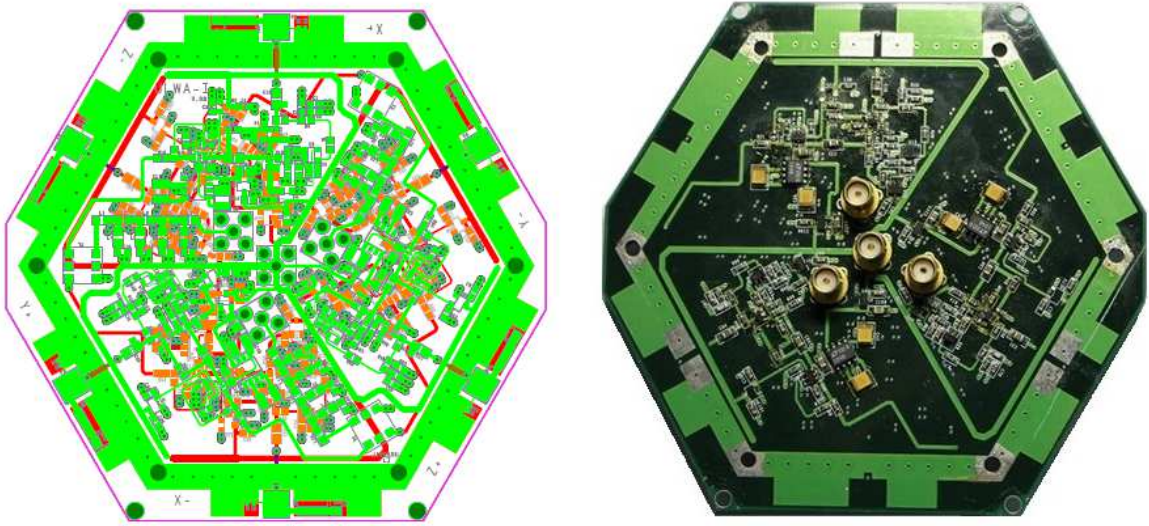}%
\caption{The prototype design of the low noise amplifiers. Left: PCB design of the LNAs. Right: Bottom view of the prototype LNAs. Note that three anti-aliasing filters are designed and combined with the LNAs on the PCB board. The four connectors are used for the power supply (central one) and the LNA outputs respectively.
\label{fig:LNADesign}}
\end{figure}

For an antenna with a very high impedance magnitude, the equivalent current noise is dominant. To make the noise of the active circuitry as small as possible, a discrete low noise Pseudomorphic High Electron Mobility Transistor (PHEMT) devices is selected to achieve the required low noise performance at low frequencies, since its high current gain and absence of base current result in a low equivalent input current. Fig.~\ref{fig:LNA} shows the circuit schematic of the low noise amplifier. Basically the LNA is designed as a two stage amplifier. The input signal is amplified by the first stage, a low noise Enhancement Mode PHEMT (E-PHEMT) amplifier ATF54143. The feedback circuit improves the stability of the amplifier. The bias voltage of the first stage is chosen to meet the noise requirements and keep the LNA stable. By increasing the bias voltage and drain current of ATF54143, the LNA performance of 1 dB compression point will be improved, as well as the second and third order output inter-modulation products (OIP2, OIP3), which means the LNA dynamic range will be increased. However, this will result in an increasing power consumption. Considering there are less strong RFIs in space, we choose to sacrifice a little dynamic range and OIP2 \(\&\) OIP3 performance to reduce the power consumption. In the final design, a trade-off has been made between OIP2 \(\&\) OIP3, 1 dB compression point and the power consumption. The output signal of the first stage is AC coupled to a current feedback amplifier (EL5162) with low power consumption. The amplified differential signals are then combined into a single-ended output by a transformer, which is optimized to match a 50-Ohm system.

\begin{figure}[t]
\centering
\setlength{\belowcaptionskip}{-0.5cm}
\includegraphics[width=\columnwidth]{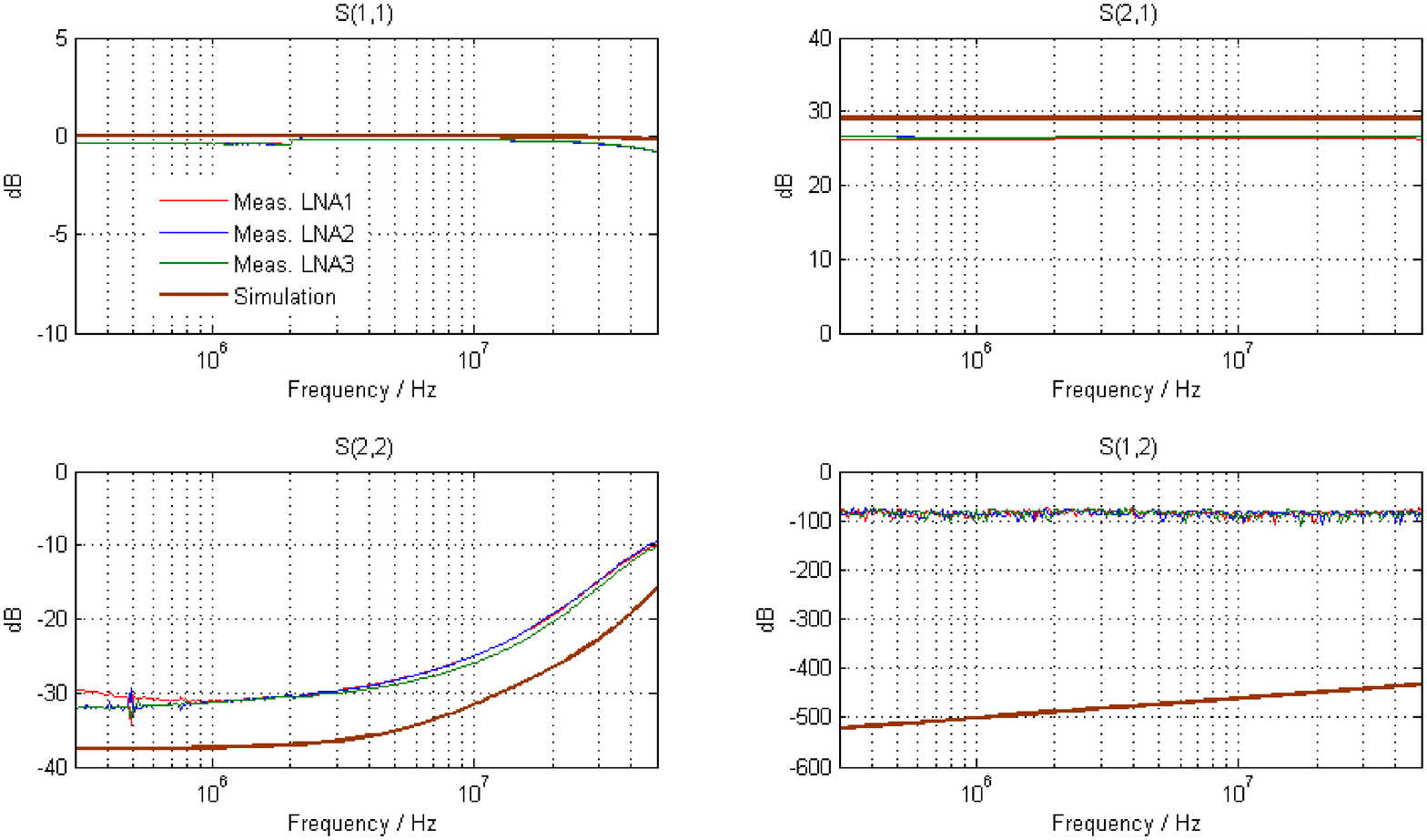}%
\caption{Simulated and measured S-parameters of the three prototype LNAs. S(1,1) and S(2,2) are defined respectively as the input and the output reflection coefficients of an network, they show how well the two-port networks match with a 50-Ohm system. S(2,1) and S(1,2) are the forward and reverse transmission gains respectively.
\label{fig:sparameter}}
\end{figure}

To evaluate the performance of a low noise amplifier several parameters can be used, such as S-parameters, stability factor, noise figure (NF), 1 dB compression point, inter-modulation, power consumption, etc. In this work all these parameters have been simulated and measured for the designed LNA. The S-parameter is normally used for the assessment of a two-port network, here both simulations and measurements of S-parameter for the LNA are provided with a reference of 50 ohms. For the simulation and measurement convenience, asymmetric signals at the LNA input are transformed into differential ones through a balun (Fig.~\ref{fig:LNADesign}).

\begin{figure}[hb]
\centering
\setlength{\belowcaptionskip}{-0.5cm}
\includegraphics[width=\columnwidth]{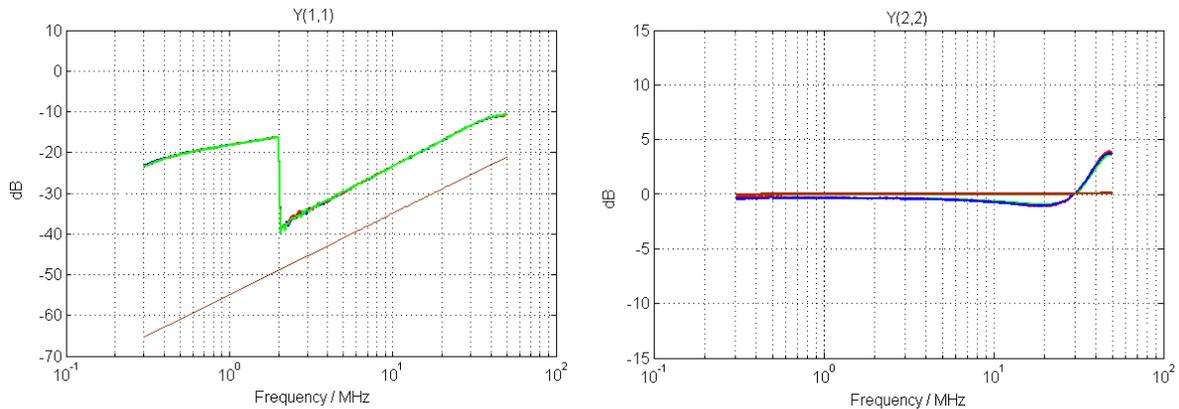}%
\caption{Simulated and measured Y-parameters of the three prototype LNAs. Y(1,1) and Y(2,2) are defined respectively as the input and the output admittance.
\label{fig:yparameter}}
\end{figure}

Figure~\ref{fig:sparameter} shows the S-parameter simulation and measurement results of the three LNA prototypes within the frequency range of 0.3 to 50 MHz. The input reflection coefficient, S(1,1) shows a good agreement between the measurements and simulations; it is close to 0 dB, which is expected for a voltage amplifier. The measurement values of the output reflection coefficient S(2,2) are higher than the simulation results over the entire band, which is caused by the non-ideal transformer characteristics at the LNA output. The measured gain S(2,1) is about 2.5 dB lower than the simulated value, in which \(0.4\sim0.5\) dB can be attributed to the insertion loss of two transformers used at the input and output ends of LNAs in the measurements. The differences between the physical characteristics of the chips used in the measurements and the chip models in the simulation also contribute to these discrepancies (about 2 dB). The simulated values of reverse gain S(1,2) are extremely low, and look like unrealistic, which could be ascribed that the amplifier models can not correctly represent the reverse isolation. Additionally, Fig.~\ref{fig:sparameter} also demonstrates well repeatability of the measured S-parameters between the the three prototype LNAs.

To understand well the properties of an input mismatching LNA described above, the Y-parameters have also been simulated and measured with the S-parameters~\cite{dean1994}. Since the forward trans-admittance Y(2,1) and backward trans-admittance Y(1,2) exhibit high consistency with the S-parameter S(2,1) and S(1,2), only the input and output admittance are plotted here as shown in Fig.~\ref{fig:yparameter}. It can be seen that the input admittance Y(1,1) is very low, which accords with the results of reflection coefficient S(1,1). This is also a typical performance for a voltage amplifier. The differences between the measurement and simulation results can be attributed to the representative differences between the models used in the simulations and the chips in the measurements. The ``break" around 2 MHz in the plot of Y(1,1) may be caused by the instrument itself (AC/DC input coupling, or band switching, etc.). Both the simulation and measurement for the output admittance Y(2,2) show good matching between the amplifier and 50 ohms.

\subsection{Sky Noise-Limited Performance}
\label{subsec:snl}For a low frequency antenna system, the sky noise-limited performance is a primary requirement. To meet this requirement, as presented in Section ~\ref{sec:temmodel}, the noise contributed by LNA must be limited to a certain level. Theoretically, we use a noise figure (NF) to describe the noise performance of a radio frequency system. It is defined as the ratio of signal-to-noise ratio (SNR) at the input port to SNR at the output port~\cite{simpson2008}.

\begin{equation}
NF=NF_{min}+\frac{4R_{n}|\Gamma_{src}-\Gamma_{opt}|^{2}}{|1+\Gamma_{opt}|^{2}(1-|\Gamma_{src}|^{2})}
\label{eqn:nf}
\end{equation}

where, \(NF_{min}\) is the minimum noise figure that the circuit can produce when the input terminal has the optimum reflection coefficient \(\Gamma_{opt}\). \(R_{n}\) is the noise resistance and it controls how fast the noise increases as the input reflection coefficient deviates from \(\Gamma_{opt}\), and \(\Gamma_{src}\) is the reflection coefficient of the input terminal. The noise temperature at the input of the LNA can be described as \(T=T_{0}(NF-1)\), where \(T_{0}\) is 290 K.

\begin{figure}[h]
\centering
\setlength{\belowcaptionskip}{-0.5cm}
\includegraphics[width=6.0in]{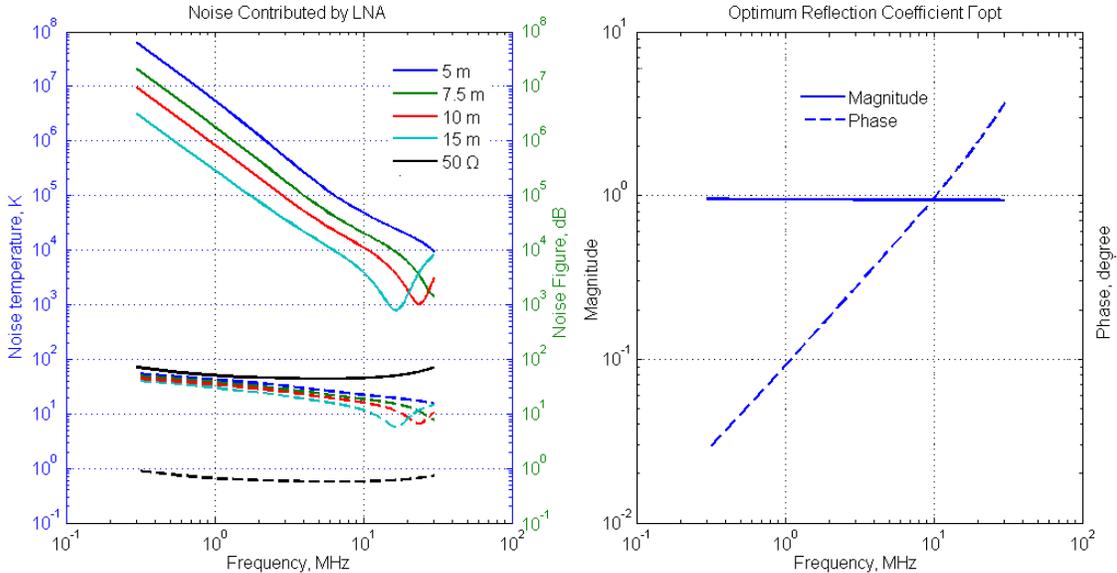}%
\caption{Left Panel: Noise contributed by the LNA connecting with tripole antennas of different lengths. 50-ohm plots are the noise of the LNA connecting with 50 ohms load. The solid lines and dashed lines denote noise temperature and noise figure, respectively. Right Panel: The optimum reflection coefficient \(\Gamma_{opt}\) of the LNA over the frequency.
\label{fig:lnanoise}}
\end{figure}

\begin{figure}[t]
\centering
\setlength{\belowcaptionskip}{-0.5cm}
\includegraphics[width=6.0in]{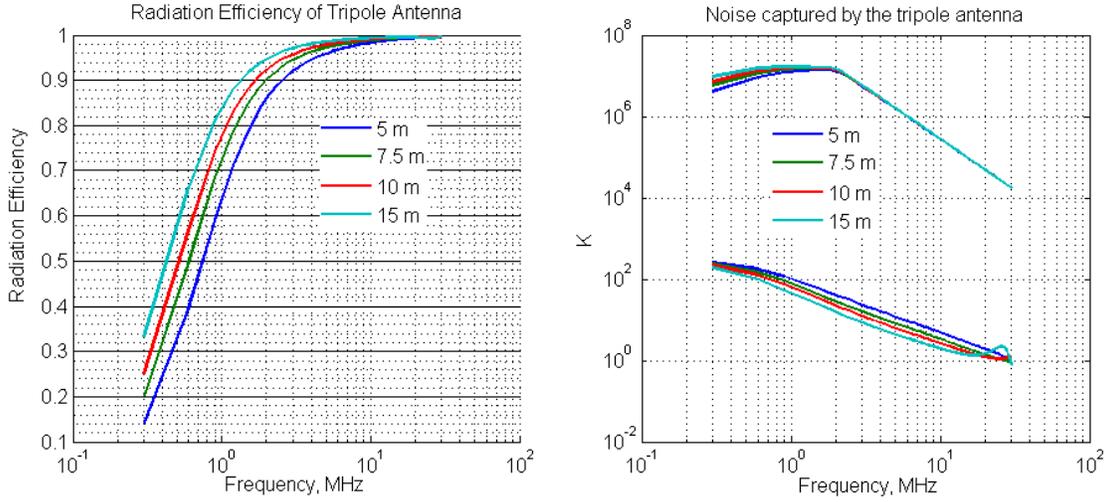}%
\caption{Left: Radiation efficiency of tripole antenna onboard the satellite. Right: Noise captured by the tripole antennas with different lengths. The upper set of curves indicate the captured sky noise temperature \(\eta T_{sky}\), and the lower set of curves show the antenna thermal noise temperature \((1-\eta)T_{0}\). \(T_{0}\) is the ambient temperature 290 k. \label{fig:antnoise}}
\end{figure}
Figure~\ref{fig:lnanoise} shows simulation results of the LNA noise connecting with the tripole antenna with different lengths. The noise temperatures appear quite high and increase toward the low end of the frequency band, which is consistent with formula (\ref{eqn:nf}). Fig.~\ref{fig:antnoise} plots the sky noise sensed by the tripole antenna, as well as the thermal noise from the antenna itself.

\begin{figure}[hb]
\centering
\setlength{\belowcaptionskip}{-0.5cm}
\includegraphics[width=\columnwidth]{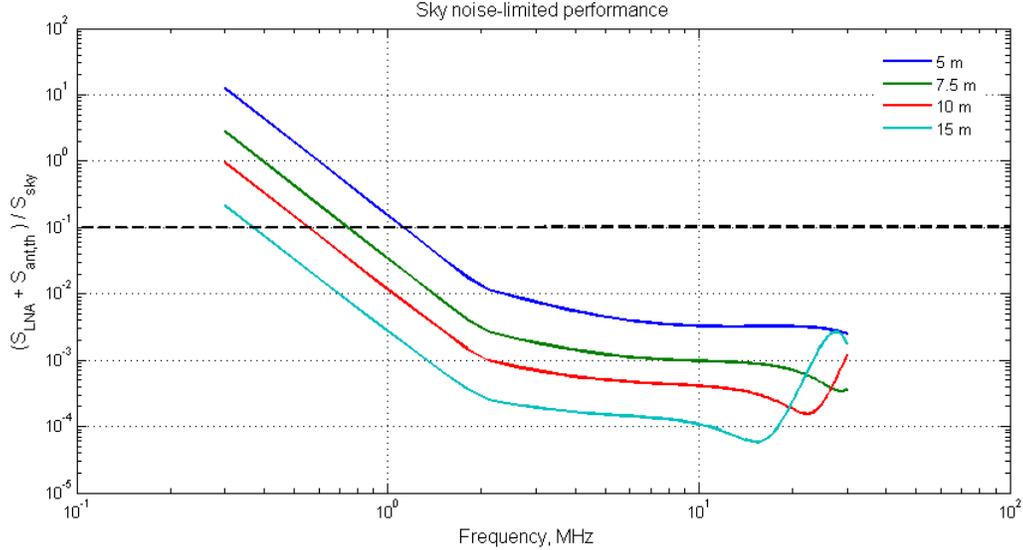}%
\caption{Sky noise-limited performance
\(\frac{S_{LNA}+S_{ant,th}}{S_{sky}}\) simulated for the tripole antennas with different lengths. The black dashed line is \(10\%\) limitation. \label{fig:skynoiselimited}}
\end{figure}

Figure~\ref{fig:skynoiselimited} shows a ratio of the receiver noise power to the sky noise power. It defines the antenna length for which the sky noise-limited operation can be achieved over the frequency range 1-30 MHz. It can be seen that for the antenna with a length of 7.5 meters, the \(10\%\) noise limitation is achievable between 0.75 MHz and 30 MHz. As for the 5-meter antenna, the \(10\%\) noise limitation is achieved between 1.1 MHz and 30 MHz. As is clear from Fig.~\ref{fig:skynoiselimited}, antenna equal or longer than 7.5 meters can achieve the sky-noise limitation performance. However, longer antenna, although offering better noise characteristics at most of the frequency band, are heavier and larger. Therefore, an optimum solution is as a short antenna as possible, in our case something near 7.5 m in length. This length is also consistent with the resonant length described in Subsec.~\ref{subsec:antenna}.

\subsection{Receiver System}
\begin{figure}[b]
\setlength{\belowcaptionskip}{-0.5cm}
\includegraphics[width=\columnwidth]{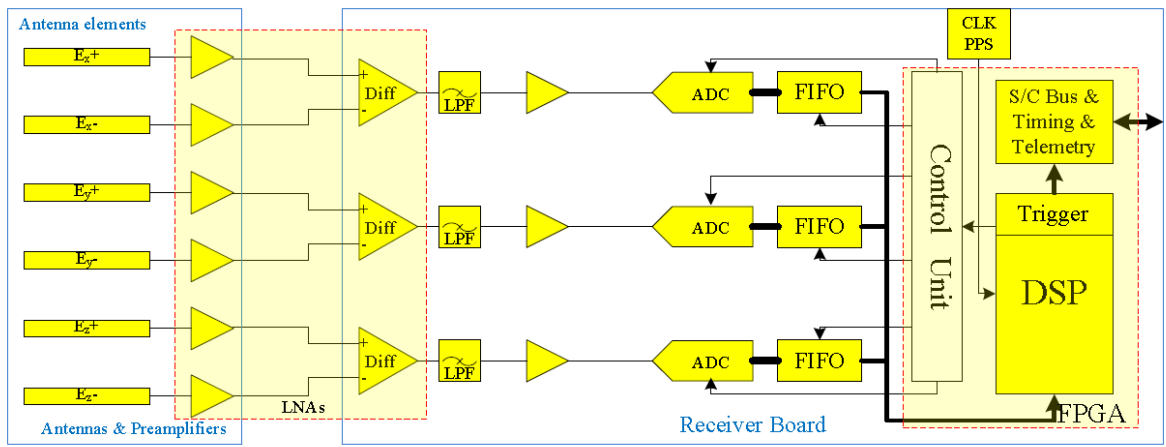}%
\caption{Block diagram of the active antenna and receiver unit.
\label{fig:systemblock}}
\end{figure}

Figure~\ref{fig:systemblock} shows the concept design of onboard ULW system consisting of the active antenna, a low pass filter and digital receiver.

The radio signals captured by the active tripole antenna first pass through an anti-aliasing low-pass filter. The low-pass filter is implemented as an Hourglass filter of 11 order with a cutoff frequency of 30 MHz and an aliasing rejection of around 55 dB at 32.5 MHz. The signals then are sampled by 14-bit Analogue-to-Digital Converters (ADC) (e.g. AD9151) operating at 65 megasamples per second (MSPS), which provides a typical Effective Number of Bits (ENOB) of 12 bits. While high sampling bits are required for the possible high dynamic range input (RFI, strong transient, etc.).The aggregate data rate for the three channels is about 340 MB/s (megabytes per second). ADCs data are fetched by high performance digital processing unit. To achieve an optimum performance of ADC, a differential driver is required to provide a flexible interface. The First Input First Output (FIFO) type of memories are employed here to buffer the sampled data from ADCs. FIFO memories can be implemented in hardware by an ultra low power SRAM. Using FIFO allows a software trigger unit to be realized in the control unit. The trigger can be utilized for transient monitoring and event selections. The control unit and digital signal processor are implemented by a FPGA with low power consumption, which controls the instrument and performs necessary pre-processing of the signals. The power dissipation of the whole receiver is about 2 Watts~\cite{mark2015}.

\begin{table}[h]
\centering
\begin{minipage}{\columnwidth}
\centering \caption{Power dissipation of antenna system and digital receiver.} \label{tab:power}
\begin{tabular}{|c|c|c|c|}
\hline
Part & Quantity & Power (mW) & Total Power (mW)\\
\hline LNA & 3 & 175\footnote{ The power dissipation of the LNA is based on laboratory measurements.}& 525\\
Receiver & 1 & \(\sim 2000\) &  \(\sim 2000\) \\
\hline
\end{tabular}
\vspace{-7pt}\renewcommand{\footnoterule}{}
\end{minipage}%
\end{table}

Table~\ref{tab:power} shows the partitions of power consumption for the antenna system. The total power dissipation is around 2.5 Watts, and is therefore consistent with the power budget limit of 4 Watts for the whole radio astronomy system.

\section{Field test with Prototype Antenna}
\label{sec:experiment}

\begin{figure}[t]
\centering
\setlength{\belowcaptionskip}{-0.5cm}
\includegraphics[width=\columnwidth]{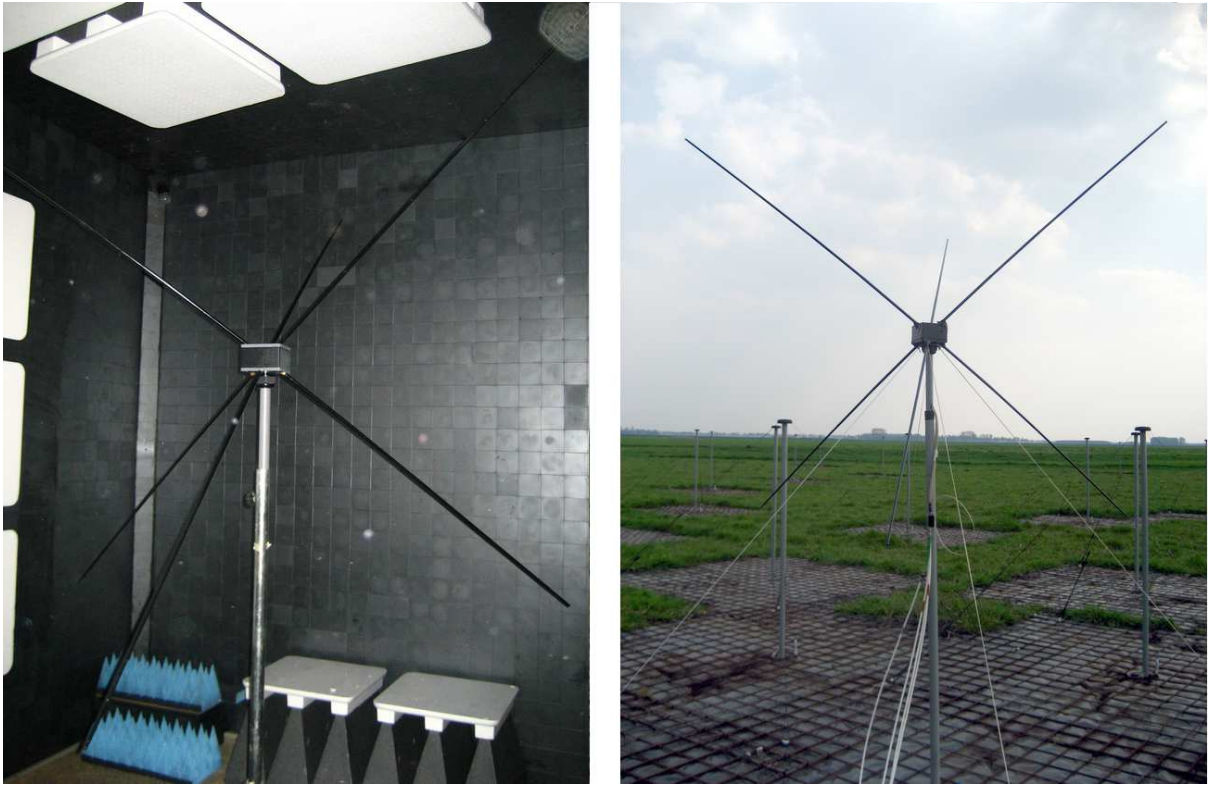}%
\caption{Tests with the prototype tripole antenna~\cite{linjie2011}. Left: Test setup in ASTRON chamber. Right: Setup in LOFAR station CS011 with 2 meters above the ground. \label{fig:prototype}}
\end{figure}

To verify the performance of the designed tripole antenna, experiments with a full-size prototype antenna are required. However, the large size of the dipole antenna (7.5 m) requires a correspondingly large anechoic test chamber which is hardly available. Therefore we conducted tests with a scaled prototype antenna smaller in size and tuned into the frequency range 30 to 87 MHz.

In~\cite{bale2007}\cite{ruck1996}, a 1/10th scale antenna model was devised for the rheometry measurements of STEREO/WAVES antenna, which are performed at high frequencies to obtain the antenna performances at corresponding low frequencies. It successfully solved the antenna size and anechoic chamber problems at low
frequencies. In order to conduct the measurements of the described tripole antenna here, we followed this scaling example by testing a prototype antenna developed originally for the LRX mission~\cite{linjie2011}\cite{marc2012} and equipped with the LNA designed in this work. This LRX prototype antenna had three 2.5-meter dipoles, and therefore could be considered as one-third scale model of the tripole antenna discussed in this paper. The scaling of the antenna results in the shift of the frequency range from \(1 - 30\) MHz to the higher frequency range \(3 - 90\) MHz. The latter range is less polluted by RFI and has enabled us to conduct the test measurements in the field, not in an anechoic chamber.

The LRX prototype antenna consists three dipoles mounted orthogonally on a cube box made of acrylonitrile-butadiene-styrene (ABS), in which the LNAs are installed and connected with the ends of dipoles (Fig.~\ref{fig:prototype}). Three cables are used to transmit the signals from LNAs to the LOFAR digital receiver, and the fourth cable is used to supply the power to LNAs. The transmission cables of LOFAR are 75-ohm cables, however, the output of proposed LNA is designed for a 50-ohm system, we use a \(\pi\) shape matching network to achieve the matching between them in the measurements. To avoid the power saturation, three 10 dB attenuators are used for the prototype antenna. All the measurements are recorded by the LOFAR ReCeiver Unit(RCU).

\begin{figure}[t]
\centering
\setlength{\belowcaptionskip}{-0.5cm}
\includegraphics[width=\columnwidth]{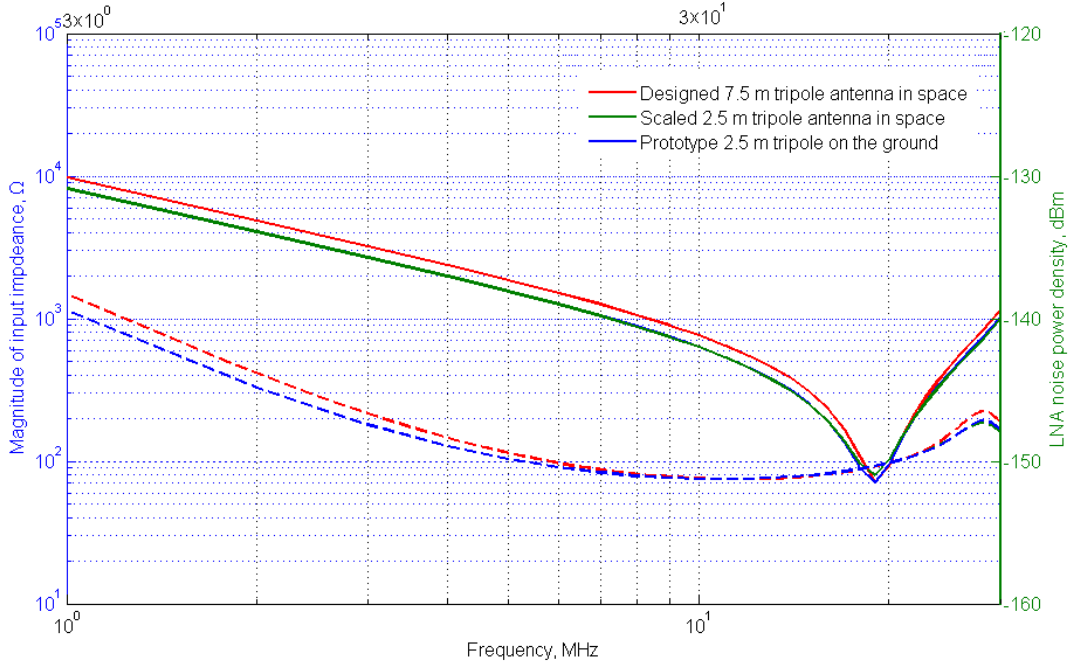}%
\caption{Simulated impedance magnitudes (solid line) and corresponding LNA noises (dashed line) for the proposed 7.5 m tripole antenna, scaled 2.5 m tripole antenna onboard the satellite, and the prototype 2.5 m tripole antenna on the ground. Here the impedance and LNA noise are simulated for 7.5 m antenna at low frequencies (lower) and for 2.5 m antenna at corresponding high frequencies (upper).
\label{fig:implna}}
\end{figure}

As defined in formula (\ref{eqn:nf}), the LNA noise is greatly influenced by the input reflection coefficient, which directly depends on the input impedance connecting with LNA. Since the field tests are carried out on the ground, we have to consider the LNA noise differences between the ground and space situations. Figure~\ref{fig:implna} shows the simulated input impedance magnitudes and LNA noises of the designed 7.5 m tripole antenna onboard satellite in space from 1 to 30 MHz, as well as the 2.5 m tripole antenna within the frequencies range of 3-90 MHz for both space (onboard satellite) and ground cases, in which the prototype 2.5 m antenna is set up 2 meters above the ground in the simulations. As shown in Fig.~\ref{fig:implna}, the results of the 2.5-meter tripole antenna are consistent well with each other for space and ground placements, and both of their impedance magnitudes and LNA noises at high frequencies agree with the results of the devised 7.5 m tripole antenna at low frequencies, especially in the frequency range from 6 to 24 MHz, which means the results of the field tests for this 2.5 m prototype antenna can be applied for evaluating the LNA noise performance of the proposed 7.5 m tripole antenna in space.

According to the analysis in Section ~\ref{sec:temmodel}, the power density at the output of the LNA can be given with formula (\ref{eqn:santt}) and (\ref{eqn:slna}) by

\begin{equation}
S_{out}=k(T_{ant} + T_{LNA})(1-|\Gamma_{ant}|^{2})G_{amp}
\end{equation}

\begin{figure}[t]
\centering
\setlength{\belowcaptionskip}{-0.5cm}
\includegraphics[width=\columnwidth]{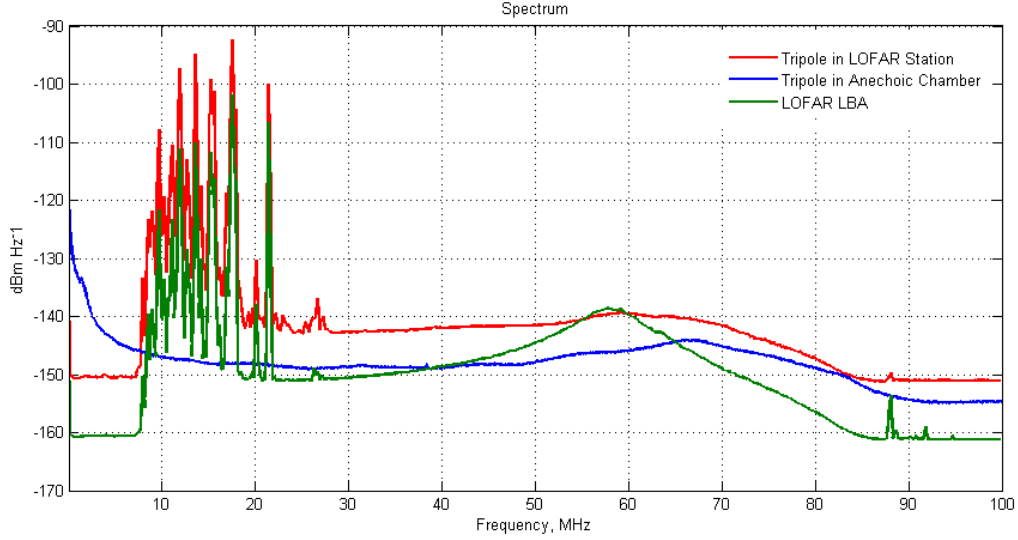}%
\caption{Spectrum measurements for the prototype antenna in EMC chamber and LOFAR station as shown in Fig.~\ref{fig:prototype}. The spectrum densities are plotted with the observation data recorded by LOFAR RCU. Note that the noise floor difference between the 2.5 m tripole antenna and LOFAR LBA results from a 10 dB attenuator used to avoid power saturation in our antenna.
\label{fig:skylofar}}
\end{figure}

To measure the LNA noise, a method similar to the Y factor method was adopted. The Y factor method is generally used for noise figure measurements~\cite{yfactor2010}. In the measurements the active antenna was installed in the microwave chamber of ASTRON and also in LOFAR central station (CS011), which provide two different noise temperatures. One is the ambient temperature in the chamber, which is almost constant over the frequency band. In the anechoic chamber, the tripole antenna only senses the ambient temperature without radio interferences. The other one is the sky noise temperature, which can be modeled with formula (\ref{eqn:skynoise}). When measured outside, the sky noise dominates the input signals for the tripole antenna, a minor influence from the ground below the antenna is expected due to the relative low temperature of the ground. Using the relative differences at the antenna output for the two noise temperatures, the LNA noise can be solved by these two measurements.

\begin{figure}[t]
\centering
\setlength{\belowcaptionskip}{-0.5cm}
\includegraphics[width=\columnwidth]{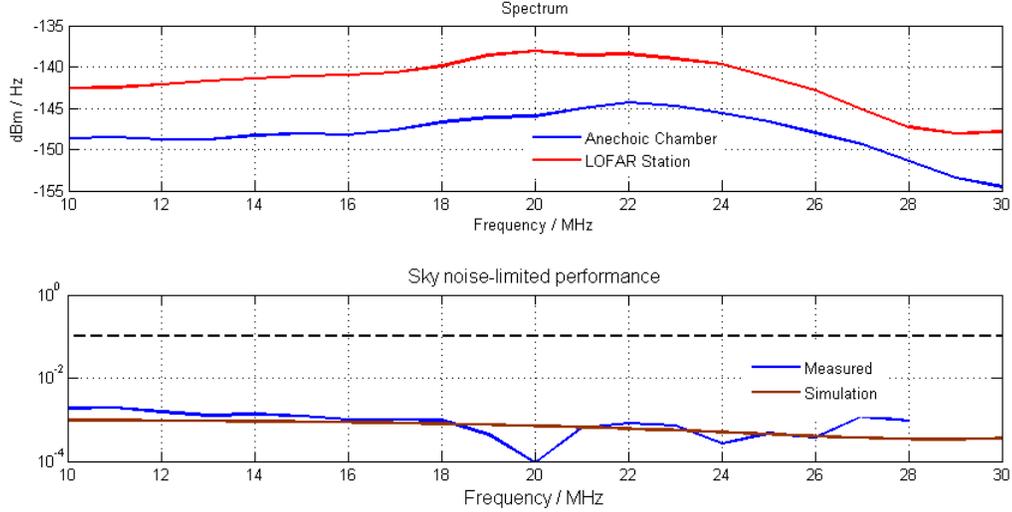}%
\caption{Sky noise-limited measurements for the prototype tripole antenna. Upper: The calibrated noise power density for the measurements in chamber and LOFAR station, the spectrum densities are plotted with the observation data recorded by LOFAR RCU. Bottom: The results of sky noise-limited performance, \(10\%\) limitation is plotted as the dashed line.
\label{fig:snlmeas}}
\end{figure}

For the two spectrum measurements of the active tripole antenna, the output noise power can be given as
\begin{eqnarray}
\label{eqn:hlpower}
\begin{cases}
P_{outH} = k(T_{ant} + T_{LNA})(1-|\Gamma_{ant}|^{2})G_{amp}B,\\[10pt]
P_{outL} = k(T_{0} + T_{LNA})(1-|\Gamma_{ant}|^{2})G_{amp}B.
\end{cases}
\end{eqnarray}
where \(P_{outH}\) and \(P_{outL}\) indicate respectively the output power recorded by the receiver for the hot noise temperature outside and the cold noise temperature inside. B represents the bandwidth of the subband. Solving for \(T_{LNA}\) gives

\begin{equation}
T_{LNA}=\frac{P_{outL}\cdot T_{ant}-P_{outH}\cdot
T_{0}}{P_{outH}-P_{outL}}
\end{equation}

Substitute \(T_{ant}\) with (1) in it, the LNA noise \(T_{LNA}\) is given as

\begin{equation}
T_{LNA}=\eta P_{outL}\frac{T_{sky}-T_{0}}{P_{outH}-P_{outL}}-T_{0}
\end{equation}

Therefore, the sky noise-limited performance can be approximated as \(\frac{T_{LNA}+(1-\eta)T_{0}}{\eta T_{sky}}\). Note that \(T_{LNA}\) here is measured at high frequencies, and is used for the calculations at corresponding low frequencies.

Figure~\ref{fig:skylofar} shows the spectrum measurements for the tripole antenna in the EMC chamber and LOFAR station. Although a 10-90MHz band-pass filter is used for the RFI suppression, due to the strong RFIs a clean power spectrum of the sky are only obtained between 30 and 87 MHz in the measurement outside of the anechoic
chamber. Considering the stop band response of the filter further, the LNA noise calculations can be merely done at frequencies below 84 MHz.

Figure~\ref{fig:snlmeas} shows the measurement result of the sky noise-limited performance between 10 and 30 MHz for the proposed 7.5 m tripole antenna using the measurements of the 2.5 m prototype antenna operating between 30 and 90 MHz. Although the LNA noise measured with the 2.5 m prototype antenna in the field is not completely representative of the LNA noise for the proposed 7.5 m tripole antenna in space, as analyzed above (Fig.~\ref{fig:implna}), we can still use it to evaluate the sky noise-limited performance for the proposed tripole antenna. From Fig.~\ref{fig:snlmeas} it can be seen that the measured results agree well with the simulation results. Below 16 MHz the measurements are a little worse than the simulations, which can be probably ascribed to the higher measured noise than the simulated noise of the LNA. Nonetheless, the measurements completely fulfill the antenna noise requirement across the measured operational frequency range. It also demonstrates the performance of the designed antenna for the space ULW radio telescope.

\section{Conclusion}
In this paper a design of an active antenna for a ULW astronomy space mission has been presented. It is shown that the 7.5 m tripole antenna in combination with a low noise amplifier and digital receiver can meet the science requirements of the DSL  or a similar ULW mission. The noise simulations have proved that the active antenna is capable of achieving the sky noise-limited performance within the frequency range \(1 - 30\) MHz. Field tests with a scaled prototype of an active antenna confirmed the noise performance between 10 MHz and 30 MHz for the 7.5 m tripole antenna. A receiver design consistent with science requirements is also discussed. The mass and power consumption estimates meet the limitations of a demonstration DSL-like mission.

\section*{Acknowledgements}
This work was supported by the grant of National Natural Science Foundation of China (NSFC) 11573043, 11203042, 11433006. LC acknowledges support by the CAS-KNAW joint PhD Training Project 07PhD10. The authors would like to thank Prof. Maohai Huang from NAOC,CAS, Mr. Michel Arts and Mr. Menno Norden from ASTRON for the help and support in this work, and the anonymous referees for many valuable comments and suggestions. The authors are grateful to ASTRON, the Netherlands, for offering access to the laboratory facilities and equipment.

\bibliographystyle{spbasic}      

\begin{thebibliography}{39}
%
%
\bibitem{yfactor2010}
Agilent Technologies,: Noise Figure Measurement Accuracy-The Y-Factor Method. Application Note 57-2, published online: https://paginas.fe.up.pt/~hmiranda/st2/an57-2.pdf
\bibitem{alexander1975}
Alexander, J. K., Kaiser, M. L., Novaco, J. C., Grena, F. R., Weber, R. R.: Scientific instrumentation of the Radio-Astronomy-Explorer-2 satellite. Astro.\&Astrophys. Vol. 40, 365-371, 1975. September 2011.
\bibitem{arts2010}
Arts, M., van der Wal, E., Boonstra, A.-J.: Antenna concepts for a space-based low-frequency radio telescope. In: ESA Antenna Workshop on Antennas for Space Applications, Noordwijk, pp. 5¨C8, 2010.
\bibitem{bale2007}
Bale, S.D.: The Electric Antennas for the STEREO/WAVES Experiment. Space Science Reviews. Vol. 136, No. 1-4, pp. 529-547, 2007.
\bibitem{burns1997}
Basart, J. P., Burns, J. O., Dennison, B. K., Weiler, K. W., Kassim, N. E., Castill, S. P., McCune, B. M.: Directions for space-based low frequency radio astronomy 1. System considerations. Radio Science. Vol. 32, NO. 1, 251~~63,January-February 1997.
\bibitem{bely1997}
Bely P. Y., Laurance R. J., Volonte S., Ambrosini R. R., Ardenne A., Barrow C. H., Bougeret J. L., Marcaide J. M., Woan G. (1997) Very Low Frequency Array on the Lunar Far Side. ESA report SCI(97)2, European Space Agency.
\bibitem{bergman2009}
Bergman, J. E. S., Blott, R. J., Forbes, A. B., Humphreys, D. A., Robinson, D. W., Stavrinidis, C.: FIRST Explorer An innovative low-cost passive formation-flying system. CEAS 2009 ¨C European Air and Space Conference. 26-29 October, 2009, Manchester, U.K.
\bibitem{baan2013}
Blott, R. J., Baan, I. W. A., Boonstra, A.-J., Bergman, J., Robinson, D., Liddle, D., Navarathinam, N., Eves, S., Bridges, C., Gao, S., Bentum, M., Forbes, A., Humphreys, D., Harroch, C.-G.: Space-based ultra-long wavelength radio observatory (low cost) - SURO-LC. European Planetary Science Congress 2013, held 8-13
September in London, UK. Online at: http://meetings.copernicus.org/epsc2013, id.EPSC2013-279.
\bibitem{dsl2016}
Boonstra, A-J., et al: Discovering the Sky at the Longest Wavelength (DSL). 2016 IEEE Aerospace Conference, Yellowstone Conference Center, Big Sky, Montana, Mar 5-12, 2016.
\bibitem{burn2012}
Burns, J. O., Lazio, J., Bale, S., Bowman, J., Bradley, R., Carilli, C., Furlanetto, S., Harker, G., Loeb, A., Pritchard, J.: Probing the first stars and black holes in the early Universe with the Dark Ages Radio Explorer (DARE). Advances in Space Research. Volume 49, Issue 3, 433-450, 2012.
\bibitem{linjie2014}
Chen, L., Zhang, M., Yan, Y., Huang, M.: The Concept of Space Ultra Long Wavelength Array. General Assembly and Scientific Symposium (URSI GASS), 2014 XXXIth URSI, 16-23 Aug. 2014, available: IEEE 10.1109/URSIGASS.2014.6929990.
\bibitem{linjie2010}
Chen, L., Aminaei, A., Falcke, H., Gurvits, L.: Optimized Estimation of the Direction of Arrival with Single Tripole Antenna. Proceeding of 2010 Loughborough Antennas\&Propagation Conference, 93-96, 8-9 November 2010, Loughborough, UK.
\bibitem{linjie2011}
Chen, L.: Research on Moon-based ultra long wavelength Radio Interferometer. Doctoral Thesis. University of Chinese Academy of Science, Beijing, November 2011.
\bibitem{compton1981}
Compton, Jr. R. T.: The tripole antenna: An adaptive array with full polarization flexibility. IEEE Trans. Antennas and Propagation. Vol. AP-29, No.6, 944-952, 1981.
\bibitem{diane1997}
Diane, F. M.: Basics of Radio Astronomy for the Goldstone-Apple Valley Radio Telescope, National Aeronautics and Space Administration Jet Propulsion Laboratory, 1997.
\bibitem{eisenberg1962}
Eisenberg, G. Z.: Short-wave antennas. Svyazizdat, Moscow, 1962.
\bibitem{ellingson2005}
Ellingson, S. W.: Antennas for the Next Generation of Low Frequency Radio Telescopes. IEEE Trans. Antennas and Propagation. Vol. 53, No. 8, August 2005, 2480-2489.
\bibitem{ellingson2009}
Ellingson, S. W., Clarke, T. E., Cohen, A., Craig, J., Kassim, N. E., Pihlstrom, Y., Rickard, L. J., Taylor, G. B.: The Long Wavelength Array. Proc. IEEE. Vol. 97, No. 8, 1421-1430, Aug. 2009.
\bibitem{bentum2010}
Engelen, S., Verhoeven, C. J. M., Bentum, M. J.: OLFAR, A Radio Telescope Based on Nano-Satellites in Moon Orbit. 24th Anuual AIAA/USU Conference on Small Satellites, Logan, UT, 2010.
\bibitem{eriksson2003}
Eriksson, S.: Study of tripole antenna arrays for space radio research. Master Thesis. Department of Astronomy and Space Physics, Uppsala University, Uppsala, Sweden, UPTEC F03 062, 2003.
\bibitem{dean1994}
Frickey, D.:Conversions Between S, Z, Y, h, ABCD, and T Parameters which are Valid for Complex Source and Load Impedances. IEEE Trans. Microwave Theory and Techniques. Vol. 42, 205 - 211, Mar. 1994.
\bibitem{herman1973}
Herman, J. R., Caruso, J. A.: Radio Astronomy Explorer (RAE)-I. Observations of terrestrial radio noise. Planet. Space Sci. Vol. 21, 443-461, 1973.
\bibitem{hicks2012}
Hicks, B. C. et al.: A wide-band, active antenna system for long wavelength radio astronomy. Publ.Astron.Soc.Pac. 124 (2012) 1090 arXiv:1210.0506 [astro-ph.IM].
\bibitem{heino2009}
Jester, S., Fackle, H.: Science with a lunar low-frequency array: From the dark ages of the Universe to nearby exoplanets. New Astronomy Review 53, 1-26, 2009.
\bibitem{kaiser1996}
Kaiser, M. L., Desch, M. D., Bougeret, J. L., Manning, R., Meetre, C. A.: WIND/WAVES observations of man-made radio transmissions. Geophysical Research Letters. Volume 23, Issue 10, 1287-1290, 1996.
\bibitem{gemmeke2009}
Kr\"{o}mer, O. et al.: New Antenna for Radio Detection of UHECR. Proceedings of the 31st International Cosmic Ray Conference, {\L}\'{o}d\'{z}, Poland (2009).
\bibitem{david2011}
Mimoun, D., et al.: Farside explorer: unique science from a mission to the farside of the Moon. Exp Astron 33: 529-585, 2011.
\bibitem{raj2015}
Rajan, R.T., Boonstra, A.-J., Bentum, M., Klein-Wolt, M., Belien, F., Arts, M., Saks, N., Veen, A. -J.:Space-based aperture array For ultra-long wavelength radio astronomy, Exp.Astron. 41 (2016) 271¨C306.
\bibitem{rhode2001}
Rhode, U. L., Whitaker, J. C.: Communications Receivers: DSP, Software Radios, and Design. 3rd ed. New York: McGraw-Hill, 2001.
\bibitem{ruck1996}
Rucker, H. O., Macher, W.,  Manning,R., Ladreiter, H. P.: Cassini model rheometry. Radio Science. Vol. 31, No. 6, 1299-1311, November-December, 1996.
\bibitem{boonstra2010}
Saks, N., Boonstra, A. J., Rajan, R. T., Bentum, M. J., Belien, F., van't Klooster, K.: DARIS, A Fleet of Passive Formation Flying Small Satellites for Low Frequency Radio Astronomy. The Small Satellites
Systems \& Services Symposium, Madeira, Portugal, May-June 2010.
\bibitem{tiehan1991}
Shen, T.: Active antenna theory and application. Xi¡¯an Jiaotong University Press (1991). ISBN£º9787560504087 / 7560504086.
\bibitem{simpson2008}
Simpson, G., Ballo, D., Dunsmore, J., Ganwani, A.: A new noise parameter measurement method results in more than 100¡Á speed Improvement and enhanced measurement accuracy. Proceeding of the 72nd ARFTG Microwave Measurement Conference, Dec 2008.
\bibitem{tan2000}
Tan, G. H., Rohner, C.: Low-frequency array active-antenna system. Proc. SPIE. vol. 4015, 446-457, Jul. 2000.
\bibitem{tingay2012}
Tingay, S. J., et al.: The Murchison Widefield Array: the Square Kilometre Array Precursor at low radio frequencies. Instrumentation and Methods for Astrophysics. 2012, available: arXiv preprint
arXiv:1206.6945.
\bibitem{haarlem2013}
van Haarlem, M. P., et al.: LOFAR: The LOw-Frequency ARray. Astronomy \& Astrophysics. Vol. 556, No. 2, Aug. 2013.
\bibitem{mark2015}
Wieczorek, M.: A mission to the farside of the Moon. A proposal in response to the call for a medium-sized mission opportunity in ESA¡¯s science programme for a launch in 2025 (M4), published online, http://www.ipgp.fr/~wieczor/MyPapers/Farside-M4-FINAL.pdf.
\bibitem{marc2012}
Klein-Wolt, M., Aminaei A., Zarka, P., Schrader, J. R., Boonstra, A. J., Falcke, H.: Radio astronomy with the Lunar Lander: opening up the last unexplored frequency regime. Planet. Space Sci. 74,
167-178, 2012.
\bibitem{zarka2012}
Zarka P., Bougeret, J. L., Briand, C., Cecconi, B., Falck, H., Girard, J., Grie{\ss}eier, J. M., Hess, S.,  Wolt, M. K., Konovalenko, A., Lamya, L., Mimoun, D., Aminaei, A.: Planetary and Exoplanetary Low Frequency Radio Observations from the Moon. Planetary and Space Science, in the special issue SPME, Volume 74, Issue 1, 156-166, 2012. doi:10.1016/j.pss.2012.08.004.

\end{thebibliography}


\end{document}